\documentclass[final,3p,times,numbers,sort&compress]{elsarticle}   

\usepackage{amsmath,amssymb,stmaryrd,amstext,amsthm}
\usepackage{mathtools}
\usepackage{bm}	
\usepackage{caption}
\usepackage{subcaption}
\usepackage{siunitx}
\usepackage{booktabs}
\usepackage[english]{cleveref}
\usepackage{lineno}
\usepackage{float}
\usepackage{graphicx}
\usepackage{verbatim}   
\newcommand{\volsym}{\rlap{\kern.08em--}V} 

\usepackage{verbatim}

\newcommand{\blue}{\color{black}} 
\newcommand{\red}{\color{black}}
\newcommand{\black}{\color{black}}

\def\tsc#1{\csdef{#1}{\textsc{\lowercase{#1}}\xspace}}
\tsc{WGM}
\tsc{QE}
\tsc{EP}
\tsc{PMS}
\tsc{BEC}
\tsc{DE}

\journal{Renewable Energy}

\begin{document}

\begin{frontmatter}



\title{An extended $k-\varepsilon$ model for wake-flow simulation of wind farms}
\author{Navid Zehtabiyan-Rezaie}
\ead {zehtabiyan@mpe.au.dk}
\author{Mahdi Abkar\corref{cor1}}
\cortext[cor1]{Corresponding author}
\ead {abkar@mpe.au.dk}

\address{Department of Mechanical and Production Engineering, Aarhus University, 8200 Aarhus N, Denmark}

\begin{abstract}
The Reynolds-averaged Navier-Stokes approach coupled with the standard $k-\varepsilon$ model is widely utilized for wind-energy applications. However, it has been shown that the standard $k-\varepsilon$ model overestimates the turbulence intensity in the wake region and, consequently, overpredicts the power output of the waked turbines. 
This study focuses on the development of an extended $k-\varepsilon$ model by incorporating an additional term in the turbulent kinetic energy equation. This term accounts for the influence of turbine-induced forces, and its formulation is derived through an analytical approach.  
To assess the effectiveness of the proposed model, we begin by analyzing the evolution of normalized velocity deficit and turbulence intensity in the wake region, and the normalized power of the waked turbines. This investigation involves a comparison of the predictions against results from large-eddy simulations in three validation cases with different layouts. We then simulate a wind farm consisting of 30 wind turbines and conduct a comparative analysis between the model-predicted normalized streamwise velocity and wind-tunnel measurements. Finally, to conclude our assessment of the proposed model, we apply it to the operational wind farm of Horns Rev 1 and evaluate the obtained normalized power with the results from large-eddy simulations. The comparisons and validations conducted in this study prove the superior performance of the extended $k-\varepsilon$ model compared to the standard version.
\end{abstract}

\begin{keyword}
Wind-farm modeling \sep
Turbine wakes \sep
Power losses \sep
Turbulence modeling  \sep
Reynolds-averaged simulation

\end{keyword}

\end{frontmatter}

\section{Introduction} \label{Sec:Introduction}
Harvesting renewable energy from the atmospheric boundary layer (ABL) through wind turbines is one of the most promising methods for clean power production. Following the policies for the mitigation of global warming, wind will be the prominent power source by 2050, generating one-third of the total electricity demand. This translates into scaling up the annual wind-energy installations by four times in the upcoming decade \cite{IRENAREP, papadis2020challenges, GLOBALWINDREPORT2022}. Such a deployment requires understanding and accurate prediction of the multi-scale fully-coupled interactions between the ABL and wind farms. Within wind farms, the turbulent flows - known as the wakes - that form behind turbines are associated with a strong velocity deficit and elevated turbulence level compared to the upstream flow. The velocity deficit decreases with distance owing to the turbulent mixing mechanism. However, due to the short inter-turbine spacing in wind farms, the velocity does not completely recover to the upstream value. This results in a power loss of the waked turbines and an increase in their structural loads (see the review of Refs. \cite{Vermeer2003, Stevens2017, PorteAgel2019Review} and references therein). Therefore, accurate predictions of turbine wakes and power losses are essential for the design of new wind farms and the optimization of the annual energy production and loads. 

Owing to the growth of computing power, computational fluid dynamics (CFD) is broadly used for wind-farm modeling.
CFD techniques, i.e., Reynolds-averaged Navier-Stokes (RANS) models \cite{Sanderse2011} and large-eddy simulation (LES) \cite{Mehta2014}, can capture the physics of flow within wind farms to a much higher extent compared to the simple analytical-empirical wake models \cite{goccmen2016wind, Archer2018,zehtab2023_note}. LES provides a high level of physical fidelity and compares well with data from experiments and field measurements \cite{Wu2013,VanderLaan2019}, nevertheless, it is costly and more suitable for problems with small scale in fundamental research. Thus, the CFD RANS approach is the preferred choice to study the interaction of ABL and wind farms in industrial applications. In particular, RANS with a $k-\varepsilon$ model is the mainstream tool for this purpose because of its low computational cost, simplicity, and high stability. However, its inherent assumptions lead to falling short of accurately predicting the turbulent flow within wind farms. More specifically, the standard $k-\varepsilon$ model overestimates the wake recovery and, consequently, overpredicts the output power of waked turbines \cite{Rethore2009,Cabezon2011,VanDerLaan2015,HornshojMoller2021}. 

Among methods utilized to compensate for the overestimation of the wake recovery by the standard $k-\varepsilon$ model, a common approach, among others, is to introduce additional terms into the transport equations of turbulence quantities. Following this approach, El Kasmi and Masson \cite{ElKasmi2008} included an extra source term in the dissipation equation as $S_\varepsilon = C_\varepsilon{}_4 \mathcal{P}/\rho k$, where $k$ denoted the turbulent kinetic energy (TKE), and $\mathcal{P}$ and $\varepsilon$ represented the production and the dissipation rates of TKE, respectively. The parameter $C_\varepsilon{}_4$ was a model constant set to 0.37, and $\rho$ was the fluid's density. The source term was active in a cylindrical region encompassing the turbine with the same cross-section as the rotor in the range $-0.25\le x/d_0\le0.25$, where $x$ and $d_0$ denoted the streamwise distance from the rotor and the rotor diameter, respectively. The velocity-deficit and turbulence intensity profiles in the wake of a single turbine obtained from their model showed a good agreement with the experimental data. However, calibration of the model coefficient and the length of the region of activity were not among the objectives of their study. 
In another work, Ren \textit{et al.} \cite{Ren2019}, inspired by the turbulent flow through tree canopies, added a source term to the TKE equation as $S_k \propto (\beta_\text{p} \overline{u}^2 - \beta_\text{d} k)$, in which $\overline{u}$ denoted mean streamwise velocity, $\beta_\text{p}$ was correlated with turbine's induction factor, and $\beta_\text{d}$ was set equal to one. For the dissipation equation, a similar model as that in Ref. \cite{ElKasmi2008} was utilized but with a parabolic model coefficient as $C_\varepsilon{}_4 = 2(2r/d_0 - 0.5)^2 + 0.2$, where $r$ represented the radial distance from the center of the rotor. The source terms were applied to a cylindrical zone with the same diameter as the rotor in the range $-0.5\le x/d_0\le0.25$. The performance of their model was investigated by applying it to the Nibe wind farm. The results were compared with field measurements, LES, and the standard $k-\varepsilon$ model. The authors reported a high dependency of the model's performance on its parameters. 
In a recent study, Li \textit{et al.} \cite{Li2022} utilized the model proposed in Ref. \cite{ElKasmi2008}, and explored the effect of $C_\varepsilon{}_4$ on the accuracy of predictions by applying it to the Horns Rev 1 (HR1), Nysted, and Wieringermeer wind farms. For the cases considered in their study, $C_\varepsilon{}_4= 0.15$ led to a good agreement between the predictions and the experimental data.
\red More recently, data-driven techniques have also been utilized to correct 
TKE in the simulation of wind-turbine wakes (e.g., see Refs. \cite{steiner2022classifying,Steiner2022}).
However, this approach requires a large amount of data and exhibits high complexity compared to the empirical models \black.

The corrective terms introduced in the above-cited works are based on empirical assumptions, and the parameters included in the models usually need tuning for new cases. 
With \red these \black in mind, we aim to develop and validate an extended $k-\varepsilon$ model with the minimum need for parameter tuning and utilization of empirical relations. 
To this end, following an analytical approach, we derive a physics-originated term associated with the impact of turbine-induced forces in the transport equation of TKE. We initiate examining the proposed model by applying it to three validation cases and comparing the results with predictions from LESs performed in-house. Next, we study a $10 \times 3$ array of wind turbines using the extended $k-\varepsilon$ model and compare the results with measurements obtained from wind-tunnel experiments. Finally, we assess the proposed model by studying the operational HR1 wind farm and evaluating the predicted normalized power against available LES data. The rest of this paper is organized as follows: In section \ref{Sec:Method}, the methodology of the study is introduced. In section \ref{Sec:Results}, the performance of the extended $k-\varepsilon$ model on different wind-farm cases is compared with LES and wind-tunnel measurements as well as the standard $k-\varepsilon$ model. Finally, the key conclusions from this study are summarized in section \ref{Sec:Conclusions}.
\section{Methodology} \label{Sec:Method}
In this section, first, the RANS framework is presented, and the additional term of the TKE equation corresponding to the turbine-induced forces is parameterized analytically. Then, details of three validation cases are given, followed by a description of our numerical setups.

\subsection{Reynolds-averaged Navier-Stokes (RANS) framework} 
The Reynolds-averaged governing equations of a turbulent incompressible flow, i.e., the conservation of mass and momentum equations, can be formulated as

\begin{equation} \label{eq:Mass}
\partial_{i} \overline{u}_{i} = 0,
\end{equation}
\begin{equation} \label{eq:Momentum}
\partial_{t} \overline{u}_{i} +\overline{u}_{j} \partial_{j} \overline{u}_{i} = -\frac{1}{\rho } {\partial_{i} \overline{p}} + {\partial_{j}} \left( 2 \nu {S}_{ij} - {R}_{ij}\right)  + f_{i}, 
\end{equation}

\noindent  where $\overline{u}_{i}$, $\overline{p}$, ${S}_{ij}$, ${R}_{ij}$, and $f_{i}$ denote the mean velocity, mean pressure, mean rate-of-strain, the Reynolds stress, and turbine-induced forces, respectively. The fluid's density and kinematic viscosity are represented with $\rho$ and $\nu$, respectively. The Reynolds-stress term is formulated through the eddy-viscosity hypothesis, i.e., ${R}_{ij}=-2 \nu_\text{T} {S}_{ij} + 2/3 k \delta_{ij}$, where, $\nu_\text{T}$ and $\delta_{ij}$ are the eddy viscosity and the Kronecker delta, respectively \cite{Pope2000,Xiao2019}. 

To model $\nu_\text{T}$, an ensemble of eddy-viscosity models is available in the literature (see, e.g., Refs. \cite{Jones1972, Launder1974, Menter1993, Shih1995, Yakhot1998}). Here, we focus on the widely used standard $k-\varepsilon$ model in which the eddy viscosity is formulated as $\nu_\text{T} = C_\mu k^2 /\varepsilon$, where $C_\mu$ is a model constant. The transport equations of $k$ in our extended $k-\varepsilon$ model is
\begin{equation} \label{eq:tke}
\partial_{t} k + \overline{u}_{i} \partial_{i} k = \partial_{i} \left[\left(\nu + \frac{\nu_\text{T}}{\sigma_k}\right) \partial_{i} k\right] + \mathcal{P} - \varepsilon + S_{k}, 
\end{equation}

\noindent where $\sigma_k$ is a constant. The additional term ($S_k$) which is absent in the standard $k-\varepsilon$ model, accounts for the impact of turbine forces that needs to be parameterized. 

\subsubsection{Parametrization of the turbine-induced terms in the RANS framework} 
We use the standard actuator-disk model without rotation to calculate the turbine-induced forces $(f_{i})$ based on the disk-averaged velocity ($\overline{u}_{\text{d},i}$) by using the disk-based thrust coefficient $C^\prime_T =C_T /(1-a)^2$, in which $a$ is the turbine's induction factor \cite{calaf2010large,HornshojMoller2021}. The formula to calculate  $f_{i}$ is

\begin{equation} \label{eq:f_u}
f_{i} = -\frac{1}{2} C^\prime_T A_\text{cell} \left(\overline{u}_{\text{d},i} n_i\right)^2 \frac{\gamma_{j,l}}{V_\text{cell}}, 
\end{equation}

\noindent where $n_i$ denotes the unit vector perpendicular to the disk. The distribution of the forces is proportional to the frontal surface of the computational mesh ($A_\text{cell})$ that falls within the rotor area. Here, $\gamma_{j,l}$ is the fraction of the area overlap between the cell at a grid point $(j,l)$ and the rotor. It has a value of 1 and 0 for all cells with and without overlap with the rotor, respectively. For cells having a partial overlap with the rotor, $\gamma_{j,l}$ is equal to the fraction of area overlap  \cite{calaf2010large}.

The transport equation of $k$ is derived through a product of fluctuating velocity by the subtraction of the Reynolds-averaged momentum equation from the instantaneous momentum equation, and then taking the average of the resulting equation. For a yaw-aligned rotor, assuming that the free stream wind is aligned with the $x$-direction, a sink term corresponding to the turbine-induced forces appears in the $k$ equation as

\begin{equation} \label{eq:Sk1}
S_{k} = -\frac{1}{2}  C'_T A_\text{cell} \left(2\overline{{u'^2_{\text{d},x}}} \overline{u}_{\text{d},x} + \overline{{u'^3_{\text{d},x}}} \right)\frac{\gamma_{j,l}}{V_\text{cell}}. 
\end{equation}

By assuming  $\overline{{u'^2_{\text{d},x}}} \approx 2/3 k_\text{d}$ and $\overline{{u'^3_{\text{d},x}}} \approx \left(\overline{{u'^2_{\text{d},x}}}\right)^{3/2}$ \cite{Pope2000}, $S_k$  can be further simplified as 

\begin{equation} \label{eq:Sk2}
S_{k} \approx -\frac{1}{2} C^\prime_T A_\text{cell} \left[\frac{4}{3} k_\text{d} \overline{u}_{\text{d},x} + \left(\frac{2}{3} k_\text{d}\right)^{3/2} \right]\frac{\gamma_{j,l}}{V_\text{cell}}, 
\end{equation}

\noindent where $k_\text{d}$ is the disk-averaged TKE.

\subsection{Description of validation cases and numerical setups} \label{sec:numSetup}
\begin{figure}[!ht]
	\centering
    \includegraphics[scale=1]{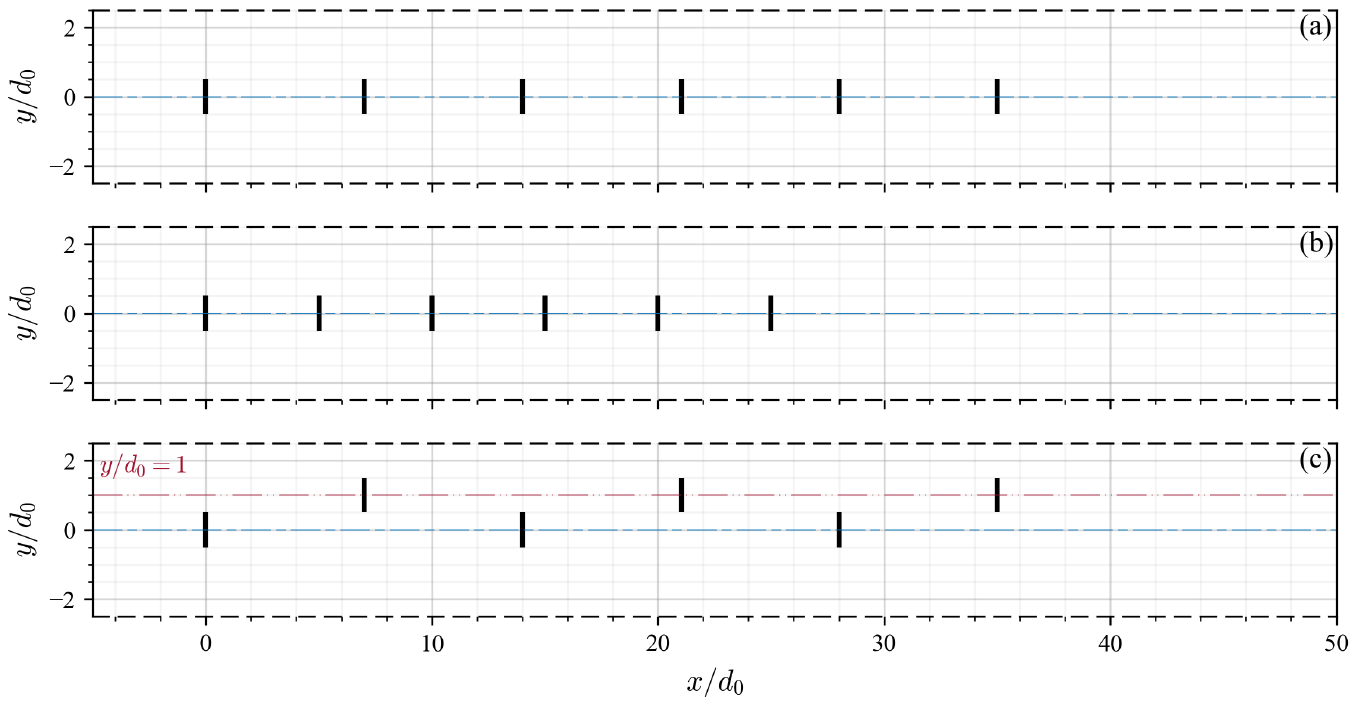}
	\caption{Schematics of the validation wind-farm cases: (a) Case 1 with a $7d_0$ streamwise spacing, (b) case 2 with a $5d_0$ streamwise spacing, and (c) case 3 with a $7d_0$ streamwise spacing where even turbines are relocated by $d_0$ in the $y$-direction. Here, the black rectangles show the turbines' location. The black dashed lines demonstrate the cyclic boundaries.}
	\label{fig:Layout_cases}
\end{figure}

Here, we define three validation cases to test the performance of the extended $k-\varepsilon$ model by comparing its predictions against those from our LESs and the standard $k-\varepsilon$ model. The validation cases have six turbines with a rotor diameter, hub height, and induction factor of 80 m, 70 m, and 0.25, respectively. The inflow velocity and inflow turbulence intensity at the hub height are 8 m/s and 5.8\%, respectively. In case 1, shown in Figure~\ref{fig:Layout_cases}(a), the inter-turbine spacing is $7d_o$ which is a typical value for wind-farm layouts \cite{meyers2012optimal}. To include a case with more severe wake effects compared to case 1, we define case 2 with an inter-turbine spacing of  $5d_o$ (Figure~\ref{fig:Layout_cases}(b)). To include a case with partial wakes, the turbines on the even rows are relocated in the $y$-direction by $d_0$ in case 3 (Figure~\ref{fig:Layout_cases}(c)). 

As our RANS framework, we use the \texttt{simpleFoam} solver, available in OpenFOAM-v2112 \cite{openfoamv2112}. We implement the new turbulence model, to solve the above-mentioned governing equations and simulate the turbine wakes and power losses under neutral ABL. The computational domains have a size of $4400\text{ m} \times 400\text{ m} \times 355\text{ m}$ in the $x$, $y$, and $z$ directions, respectively. In all three cases, the first turbine is located at a distance of $5d_0$ from the domain's inlet. Our grid-convergence study shows a resolution of $234\times40\times58$ in the $x$, $y$, and $z$ directions is sufficient for all the simulations. 
We enhance the mesh resolution in the $x$-direction at the turbines' location. We also maintain a uniform division of the domain in the spanwise direction while employing a stretching grid in the $z$-direction.
We apply the cyclic boundary conditions at the side boundaries. A pressure-outlet condition for the pressure, and a zero-gradient boundary condition for all other quantities are utilized at the outlet. At the inlet and top boundaries, for the velocity and turbulence quantities, values corresponding to a logarithmic boundary-layer flow \cite{van2019improved} are utilized, and for pressure, a zero-gradient condition is used. A rough wall boundary condition with wall functions for the turbulence quantities is imposed at the bottom wall \cite{Eidi2022}.

The LES framework and the numerical setups utilized for the simulation of the validation cases are similar to our previous works, and for the sake of brevity, we do not include further details on those. Readers are referred to Refs. \cite{Eidi2021,Eidi2022}, for more information. The LES framework has been validated and extensively used in prior studies, e.g., see Refs. \cite{porte2011large,Abkar2014,Abkar2015,Abkar2016,bastankhah2019multirotor}.
\section{Results and discussion} \label{Sec:Results}
In this section, first, a thorough analysis is conducted on the performance of the extended $k-\varepsilon$ model when applied to the validation cases. The predicted normalized velocity deficit ($\Delta \overline{u} / \overline{u}_\text{h} = (\overline{u}_\text{in} - \overline{u}) / \overline{u}_\text{h}$ with $\overline{u}_\text{in}$ and $\overline{u}_\text{h}$ as the inflow velocity and the hub-height inflow velocity, respectively) and turbulence intensity ($I = \sqrt{2k/3}/\overline{u}_\text{h}$) are compared against those from LES. \blue 
Aiming to provide a more comprehensive and quantifiable understanding of the physical phenomena under consideration, we also examine the evolution of the ratio of eddy viscosity to the inflow value ($\nu_\text{T}/\nu_{\text{T}_0}$) in case 1, predicted through the standard and extended $k-\varepsilon$ models\black.
Afterward, the vertical profiles of normalized streamwise velocity ($ \overline{u} / \overline{u}_\text{h}$) downstream of several rows in a $10 \times 3$ array of turbines are compared with the wind-tunnel measurements performed by Chamorro and Port{\'e}-Agel \cite{Chamorro2011}. Finally, we apply our model to the HR1 wind farm and evaluate its predictions against LES data from Wu and Port\'e-agel \cite{Wu2015}. The extended $k-\varepsilon$ model would be considered successful if it could outperform the standard $k-\varepsilon$ in all the mentioned cases. 

\subsection{Evaluating the extended $k-\varepsilon$ model on the validation cases}
\begin{figure}[!ht]
	\centering
	\includegraphics[scale=1]{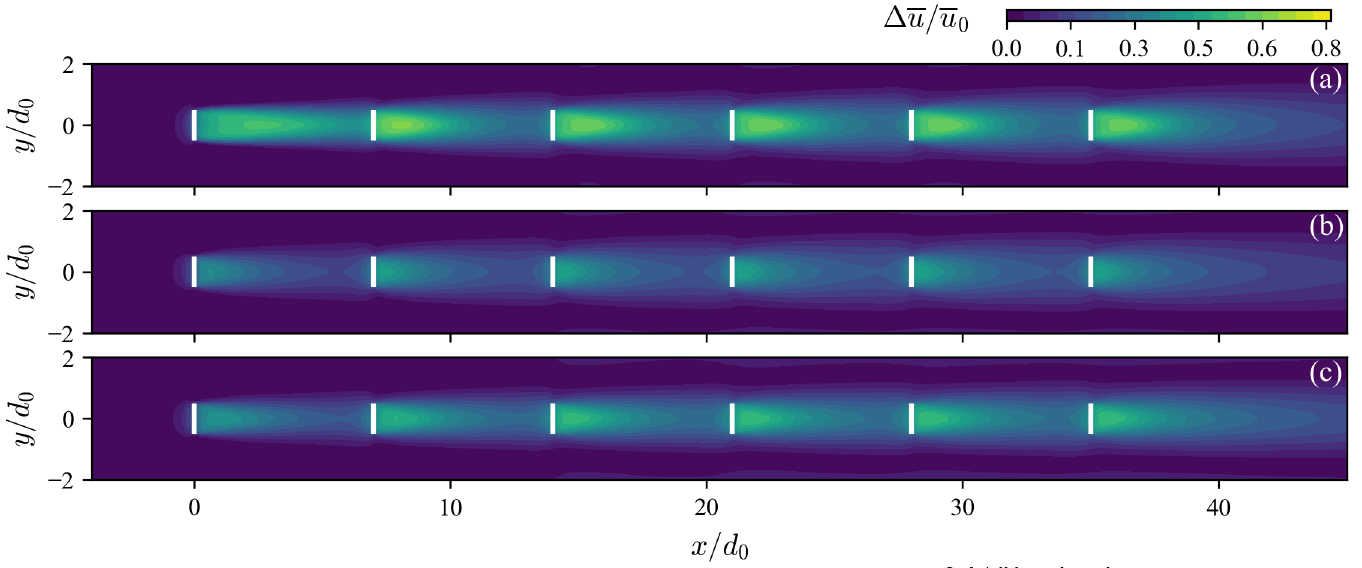}
	\caption{Distribution of the normalized velocity deficit at the hub height in case 1, predicted by (a) LES, (b) standard $k-\varepsilon$, and (c) extended $k-\varepsilon$. The white rectangles show the turbines' location.}
	\label{fig:cont_caseI_deficit}
\end{figure}

In Figure~\ref{fig:cont_caseI_deficit}, two-dimensional representations are adopted to show the normalized velocity deficit at the hub height in case 1, obtained from LES, the standard $k-\varepsilon$ model, and the extended $k-\varepsilon$ model, from top to bottom. Compared to the results from LES and the extended $k-\varepsilon$ model, the standard $k-\varepsilon$ model underestimates the velocity deficit behind all turbines, with the largest deviation observed for the most upstream one. 
It is essential to recognize that turbines exert a substantial pressure gradient upon the flow within their very vicinity. This circumstance poses a significant challenge to the eddy-viscosity models due to their inherent limitation in capturing pressure-velocity correlations \cite{Rethore2009}. Consequently, this limitation impacts the accuracy of predictions rendered by both the standard $k-\varepsilon$ model and the extended $k-\varepsilon$ model in the near wake. In this context, focusing on the downstream distribution of normalized velocity deficit, i.e., at $x/d_0 \geq 5$, shows that the extended $k-\varepsilon$ model has successfully captured a distribution close to LES which is essential for an accurate prediction of normalized power of the waked turbines.

\begin{figure}[!ht]
	\centering
        \includegraphics[scale=1]{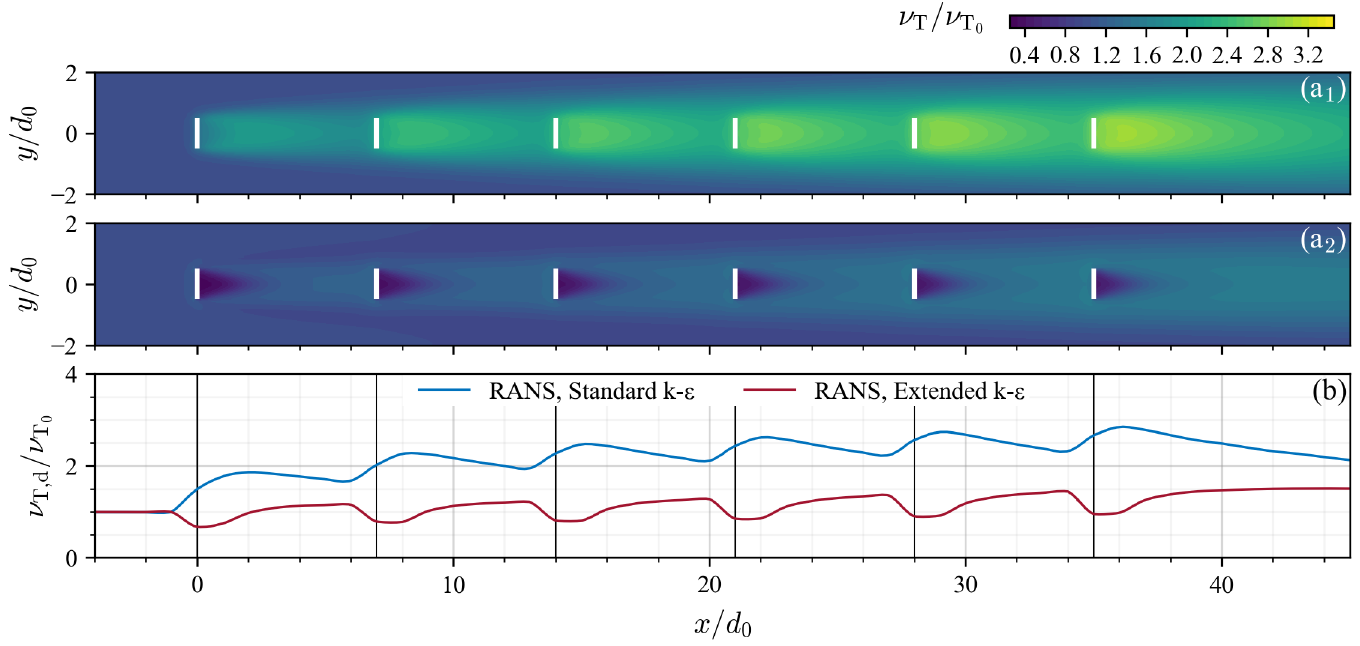}
	\caption{\blue Distribution of the ratio of eddy viscosity to the inflow value ($\nu_\text{T}/\nu_{\text{T}_0}$) at the hub height in case 1, predicted by ($\text{a}_1$) the standard $k-\varepsilon$ and ($\text{a}_2$) the extended  $k-\varepsilon$. (b) The rotor-averaged ratio of eddy viscosity to the inflow value ($\nu_\text{T,d}/\nu_{\text{T}_0}$) in case 1, predicted by the standard and extended $k-\varepsilon$ models\black.}
	\label{fig:cont_nut}
\end{figure}

\blue
To shed light on the phenomena observed in the previous figure, and to comprehend the physics of flow in turbine wakes as captured by the standard and extended $k-\varepsilon$ models, we study the evolution of eddy viscosity in case 1. Figures~\ref{fig:cont_nut}($\text{a}_1$) and ($\text{a}_2$) illustrate the hub-height level contours of the ratio of eddy viscosity to the inflow value ($\nu_\text{T}/\nu_{\text{T}_0}$). As seen in these figures, the term introduced in the TKE equation successfully addresses the well-known issue associated with the standard $k-\varepsilon$ model regarding the overestimation of the eddy viscosity and mixing \cite{Sanderse2011, HornshojMoller2021}.
Additionally, Figure~\ref{fig:cont_nut}(b) presents the rotor-averaged $\nu_\text{T}/\nu_{\text{T}_0}$, where rotor-averaging is performed over the intersection area of the $x$-normal planes and a hypothetical cylinder starting from the inlet with the same diameter as that of the turbines with its axis at the hub height. It is evident that the extended $k-\varepsilon$ model significantly reduces $\nu_\text{T}/\nu_{\text{T}_0}$ in the wake region at all $x$-positions. These findings are reasonable to extrapolate to cases 2 and 3 and, for the sake of brevity, we omit the results for these cases. 
\black

\begin{figure}[!ht]
	\centering
	\includegraphics[scale=1]{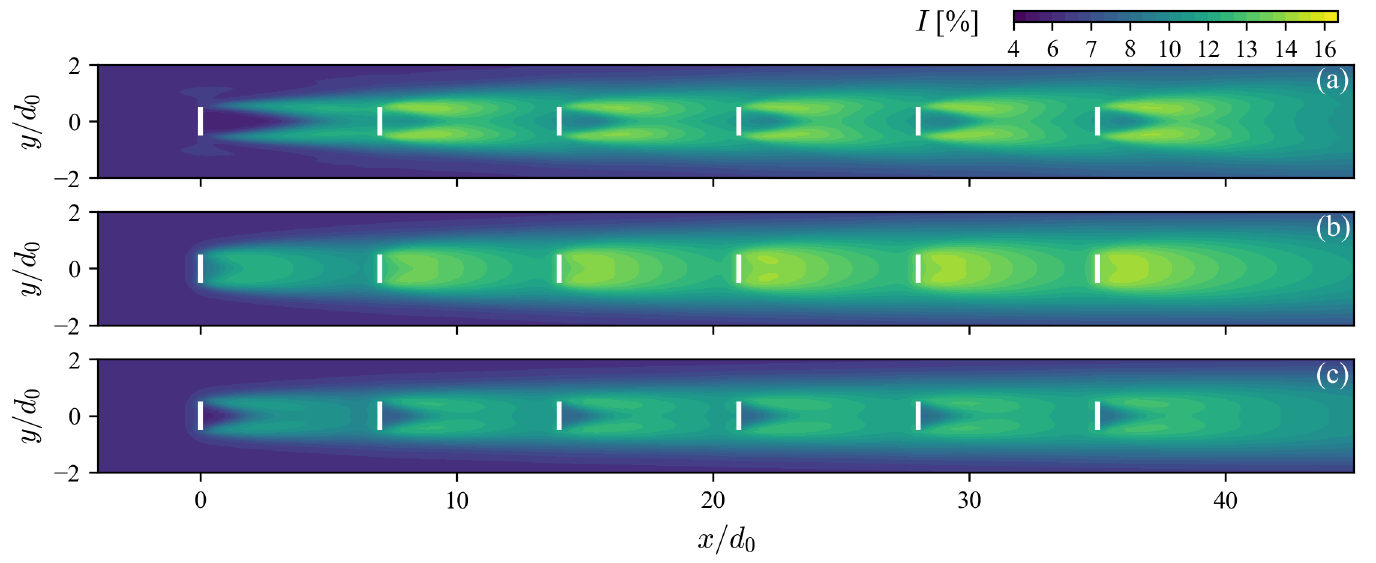}
	\caption{Distribution of the turbulence intensity at the hub height in case 1, predicted by (a) LES, (b) standard $k-\varepsilon$, and (c) extended $k-\varepsilon$.}
	\label{fig:cont_caseI_TI}
\end{figure}

Figure~\ref{fig:cont_caseI_TI} depicts contour plots of turbulence intensity at the hub height predicted through LES, the standard $k-\varepsilon$ model, and the extended $k-\varepsilon$ model. The deficiency of the standard $k-\varepsilon$ model in the prediction of the turbulence level in the wake region is visible here. Unlike LES and the extended $k-\varepsilon$ model (Figures~\ref{fig:cont_caseI_TI}(a) and (c)), the standard $k-\varepsilon$ model (Figure~\ref{fig:cont_caseI_TI}(b)) fails to capture the double peak structure of the turbulence intensity behind the turbines. 
\blue This is consistent with the distribution of eddy viscosity presented in Figure~\ref{fig:cont_nut}($\text{a}_1$), where the standard $k-\varepsilon$ model exhibited a single-peak behavior with a maximum at the center of the wake across all locations. In contrast, the extended $k-\varepsilon$ model (Figures~\ref{fig:cont_nut}($\text{a}_2$)) demonstrated an opposite behavior with a double-peak structure, leading to a reduction of $\nu_\text{T}/\nu_{\text{T}_0}$ in the central part of the wake compared to the sides\black.
Owing to the overestimation of the eddy viscosity and enhanced mixing, the standard $k-\varepsilon$ model predicts a strong development of turbulence intensity in turbine wakes, especially behind the first turbine. Turning to the extended $k-\varepsilon$ model, the new term included in the TKE equation acts as a sink at the location of the turbines and plays an important role in controlling the production of turbulent kinetic energy in the wake region.  

\begin{figure}[!ht]
	\centering
	\includegraphics[scale=1.2]{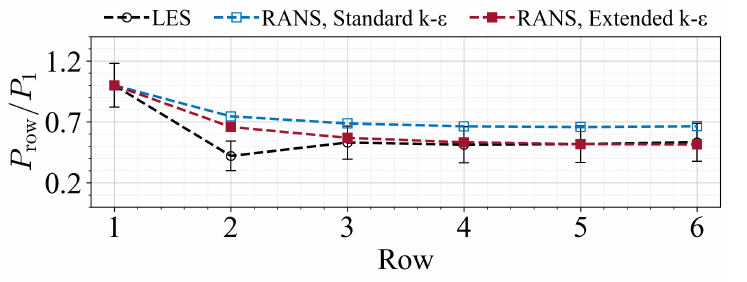}
	\caption{The normalized power as a function of turbine row in case 1. Here, normalized power is calculated by dividing the output power of each row by that of the first row through $P_\text{row}/P_1 \times 100$. The vertical bar indicates the standard deviation of the normalized power obtained from LES.}
	\label{fig:NP_case1}
\end{figure}

To assess the performance of the extended $k-\varepsilon$ model more quantitatively, the normalized power of turbines is analyzed. The success level of a model in the prediction of the normalized power is highly dependent on the accuracy of the obtained disk-averaged velocity. Therefore, with the results of the previous figures in mind, we expect the extended $k-\varepsilon$ model to outperform the standard version. Figure~\ref{fig:NP_case1} compares the turbines' normalized power predicted by the extended $k-\varepsilon$ model with the values obtained from LES and the standard $k-\varepsilon$ model. Apart from the second turbine, the extended $k-\varepsilon$ model shows a good agreement with the LES and outperforms the standard $k-\varepsilon$ model for all the waked turbines. The overprediction of the normalized power for the second turbine can be primarily attributed to a substantial velocity recovery within the wake of the most upstream turbine, which results from the strong turbulent mixing predicted through RANS models. 

\begin{figure}[!ht]
	\centering
	\includegraphics[scale=1]{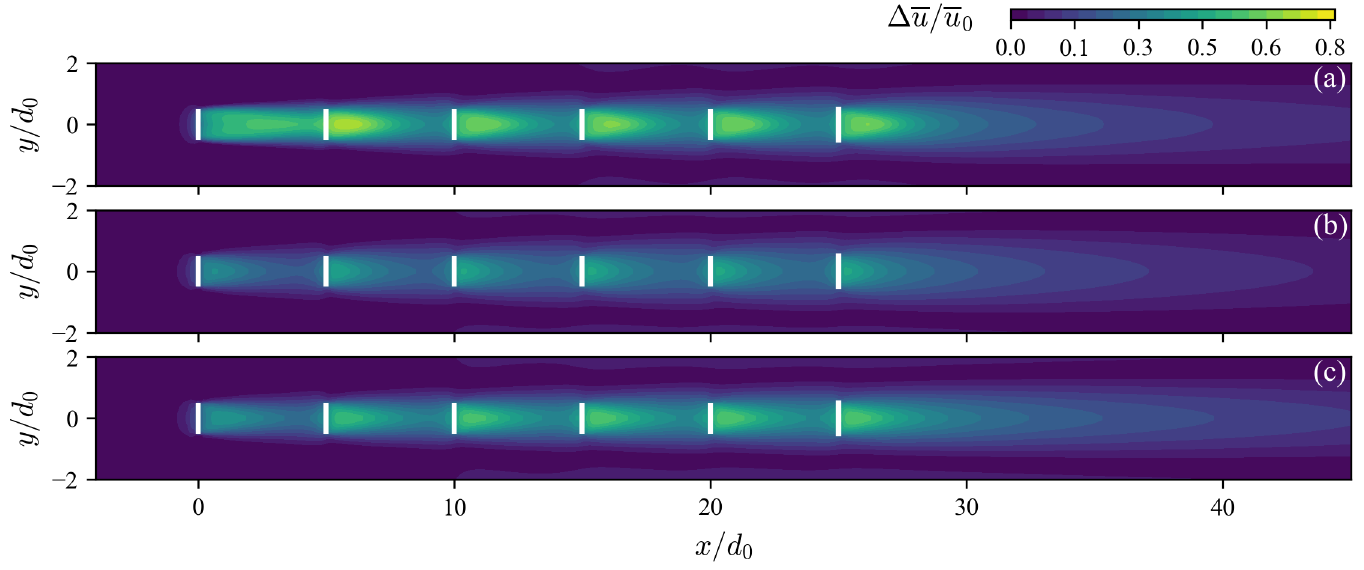}
	\caption{Distribution of the normalized velocity deficit at the hub height in case 2, predicted by (a) LES, (b) standard $k-\varepsilon$, and (c) extended $k-\varepsilon$.}
	\label{fig:cont_caseII_deficit}
\end{figure}

\begin{figure}[!ht]
	\centering
	\includegraphics[scale=1]{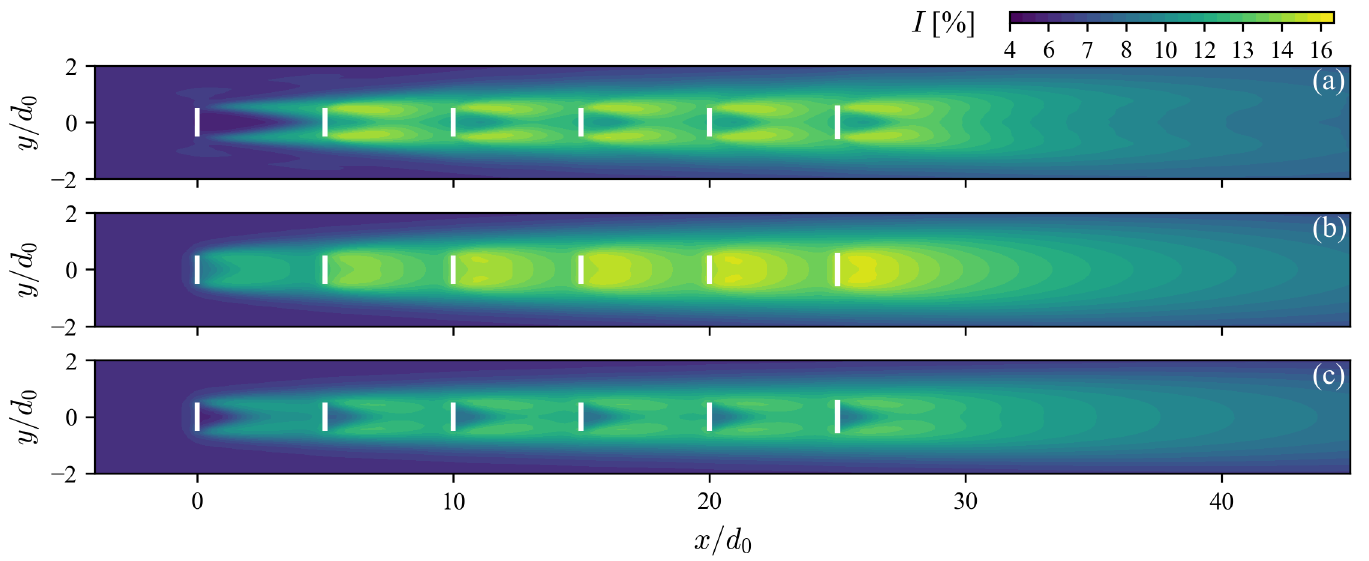}
	\caption{Distribution of the turbulence intensity at the hub height in case 2, predicted by (a) LES, (b) standard $k-\varepsilon$, and (c) extended $k-\varepsilon$.}
	\label{fig:cont_caseII_TI}
\end{figure}

Figures~\ref{fig:cont_caseII_deficit} and \ref{fig:cont_caseII_TI} depict the contour plots of normalized velocity deficit and turbulence intensity at the hub height in case 2. As mentioned earlier, this case has a stronger wake effect compared to case 1 due to a shorter inter-turbine spacing. The predicted normalized velocity deficit by the extended $k-\varepsilon$ model (Figure~\ref{fig:cont_caseII_deficit}(c)) is closer to the LES results (Figure~\ref{fig:cont_caseII_deficit}(a)), especially for the waked turbines. This shows that the improvement compared to the standard $k-\varepsilon$ model would be higher in cases with strong wake effects. Similarly, the term introduced in the TKE equation is helping to hamper the overprediction of the turbulence intensity behind the turbines. Also, the double-peak structure of the turbulence intensity behind the turbines is captured by the extended $k-\varepsilon$ model (Figure~\ref{fig:cont_caseII_TI}(c)) similar to LES (Figure~\ref{fig:cont_caseII_TI}(a)).

\begin{figure}[!ht]
	\centering
	\includegraphics[scale=1.2]{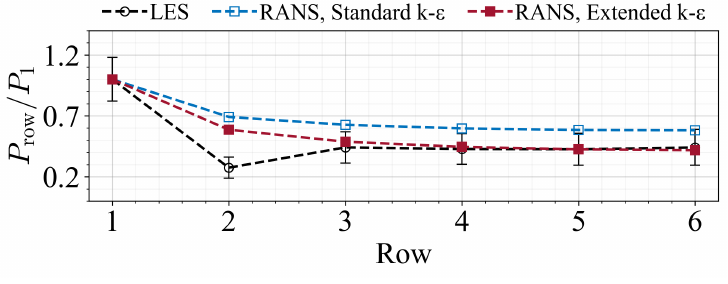}
	\caption{The normalized power as a function of turbine row in case 2.}
	\label{fig:NP_case2}
\end{figure}

The normalized power of turbines in case 2, obtained from LES, the extended $k-\varepsilon$ model, and its standard version are shown in Figure~\ref{fig:NP_case2}. Due to the higher wake losses of case 2 compared to case 1, the normalized powers of all waked turbines are degraded in this case. A good agreement between the results of LES and the extended $k-\varepsilon$ can be observed, especially for turbines 3 to 6. Much like the scenario witnessed in case 1, the strong development of RANS-predicted turbulence intensity behind the very first turbine due to overestimation of eddy viscosity and, as a result, a fast recovery of velocity, underlies the higher error of the normalized power for the second turbine.

\begin{figure}[!ht]
	\centering
	\includegraphics[scale=1]{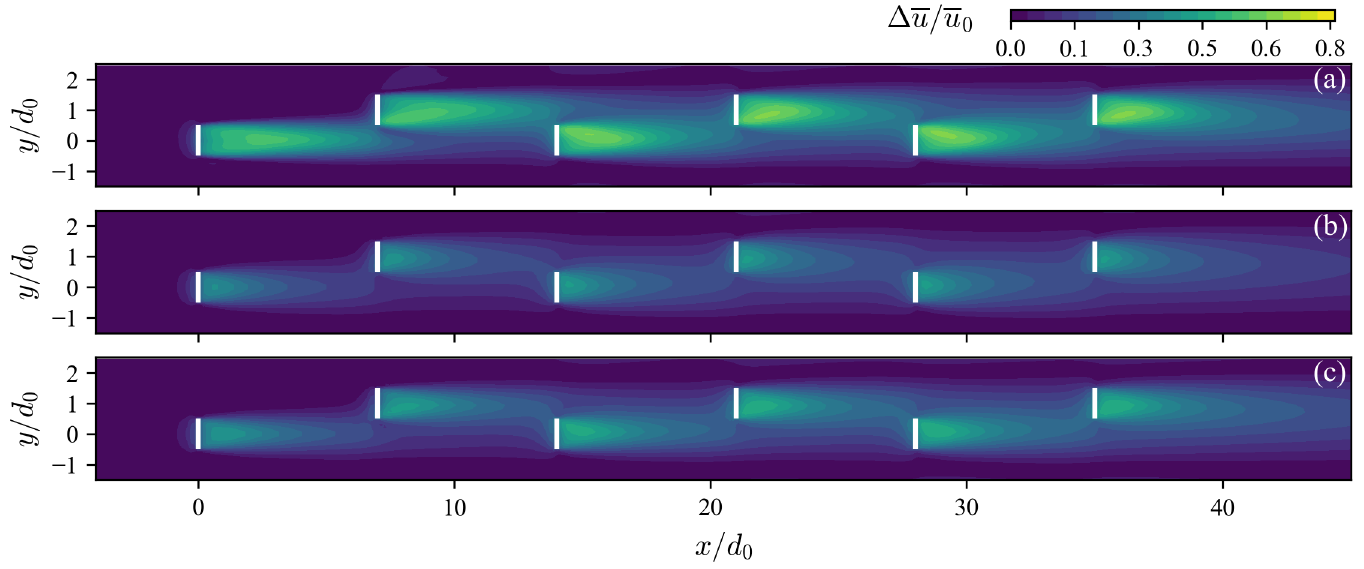}
	\caption{Distribution of the normalized velocity deficit at the hub height in case 3, predicted by (a) LES, (b) standard $k-\varepsilon$, and (c) extended $k-\varepsilon$.}
	\label{fig:cont_caseIII_deficit}
\end{figure}

\begin{figure}[!ht]
	\centering
	\includegraphics[scale=1]{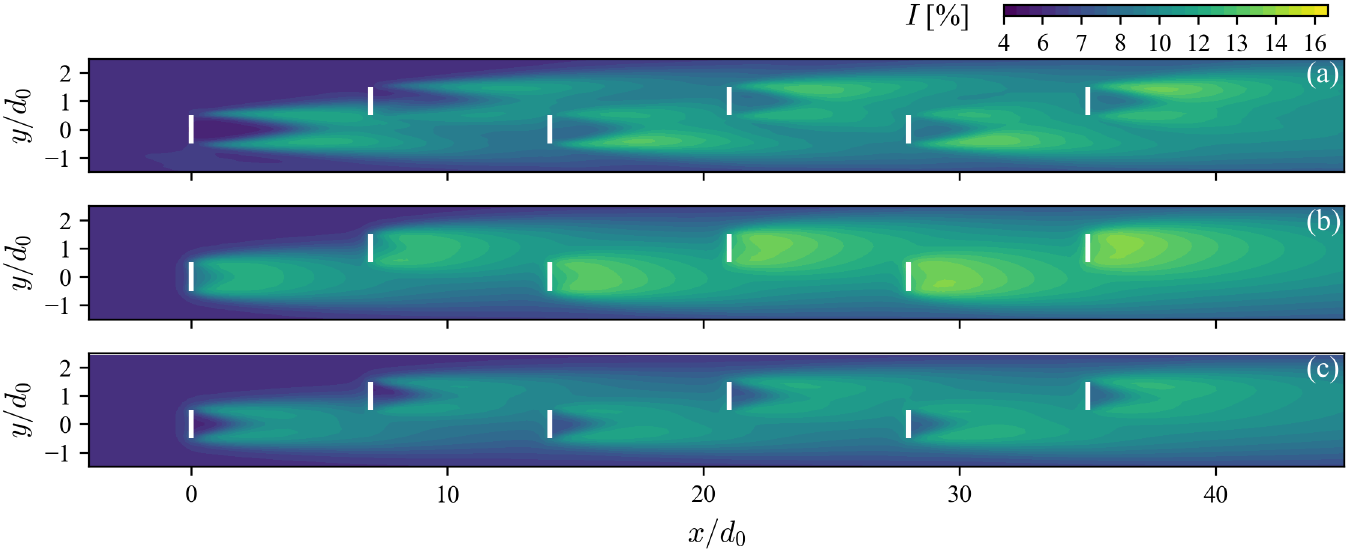}
	\caption{Distribution of the turbulence intensity at the hub height in case 3, predicted by (a) LES, (b) standard $k-\varepsilon$, and (c) extended $k-\varepsilon$.}
	\label{fig:cont_caseIII_TI}
\end{figure}

Figures~\ref{fig:cont_caseIII_deficit} and \ref{fig:cont_caseIII_TI} illustrate the hub-height contour plots of normalized velocity deficit and turbulence intensity in case 3. In this case, downstream-located turbines experience partial wakes and mild full wakes. Thus, as observed in Figure~\ref{fig:cont_caseIII_deficit}, the normalized velocity deficit is smaller than that in the cases investigated earlier. In terms of matching with LES results, in a similar manner as the previous cases, we can observe a good agreement for the extended $k-\varepsilon$ model. According to Figure~\ref{fig:cont_caseIII_TI}, and by focusing on the LES results, one can see that the added turbulence in the wake region of case 3 is also smaller compared to cases 1 and 2. The overestimation of the turbulence intensity by the standard $k-\varepsilon$ model can be observed in this figure. On the contrary, the turbulence intensity obtained from the extended $k-\varepsilon$ model shows a good agreement with the LES results. 

\begin{figure}[!ht]
	\centering
	\includegraphics[scale=1.2]{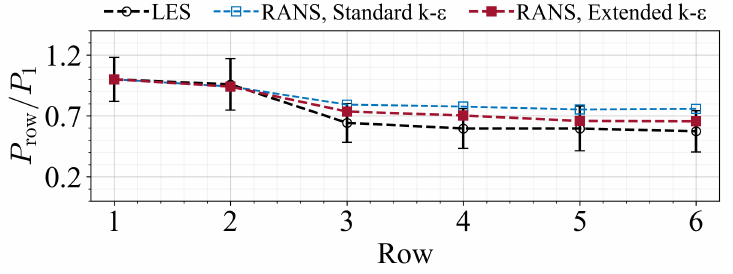}
	\caption{The normalized power as a function of turbine row in case 3.}
	\label{fig:NP_case3}
\end{figure}

Figure~\ref{fig:NP_case3} illustrates the normalized power in case 3 calculated through LES, the standard $k-\varepsilon$ model, and our extended version. Owing to the fact that turbine 2 is only affected by a partial wake from turbine 1, its normalized power is significantly higher than the second turbines in cases 1 and 2, and a perfect match between the three models is observed. Focusing on the other waked turbines, the extended $k-\varepsilon$ model performs satisfactorily, with a good agreement with the results from LES. 

Table~\ref{tab:case_summary} provides the average value of the normalized power of waked turbines (No. 2 - 6) calculated through the three models of LES, standard, and extended $k-\varepsilon$. The maximum errors compared to LES occur in case 2 with the most severe wake effect for both RANS models, however, the extended $k-\varepsilon$ model has a considerably lower error compared to the standard version. The reduction in error is maximum (37.50\%) for case 2, and in case 3 where the standard $k-\varepsilon$ model performs well, the reduction in error is 8.95\%. Therefore, we can infer that the strategy for including the analytically-derived term in the TKE equation is successful and the extended model performs well when applied to an ensemble of wind-farm cases with different layouts and levels of wake strength. 

\begin{table}[!ht]
\centering
\caption{Average normalized power of the waked turbines in cases 1-3 predicted by LES, standard $k-\varepsilon$, and extended $k-\varepsilon$. Relative errors of RANS methods compared to LES are calculated through $\lvert\text{ LES - RANS }\lvert/\text{LES} \times 100$. Error reduction is calculated by subtracting the error of the extended $k-\varepsilon$ from that of the standard $k-\varepsilon$.}
\label{tab:case_summary}
\begin{tabular}{@{}c|ccc|cc|c@{}}
\toprule
 Case  &	 \multicolumn{3}{c|}{Average of the normalized power of waked turbines}    &	\multicolumn{2}{c|}{Error}          &   Error\\
No.      &	LES	             &	Standard $k-\varepsilon$   &	Extended $k-\varepsilon$ & Standard $k-\varepsilon$  &	Extended $k-\varepsilon $          &  reduction  \\ \midrule
Case 1	   &	0.50	     &	0.68	 & 0.56 &         	36.00\%			       &	12.00\%		&    24.00\%	     \\	
Case 2	   &	0.40	     &	0.62	  & 0.47&         	55.00\%			       &	17.50\%		 &   37.50\%	     \\	  
Case 3	   &	0.67	     &	0.80	  &0.74&         	19.40\%			       &	10.45\%		  & 8.95\%	     \\			\bottomrule
\end{tabular}
\end{table}

\subsection{Evaluating the extended $k-\varepsilon$ model against wind-tunnel experiments}

\begin{figure}[!ht]
\centering
\includegraphics[scale=1]{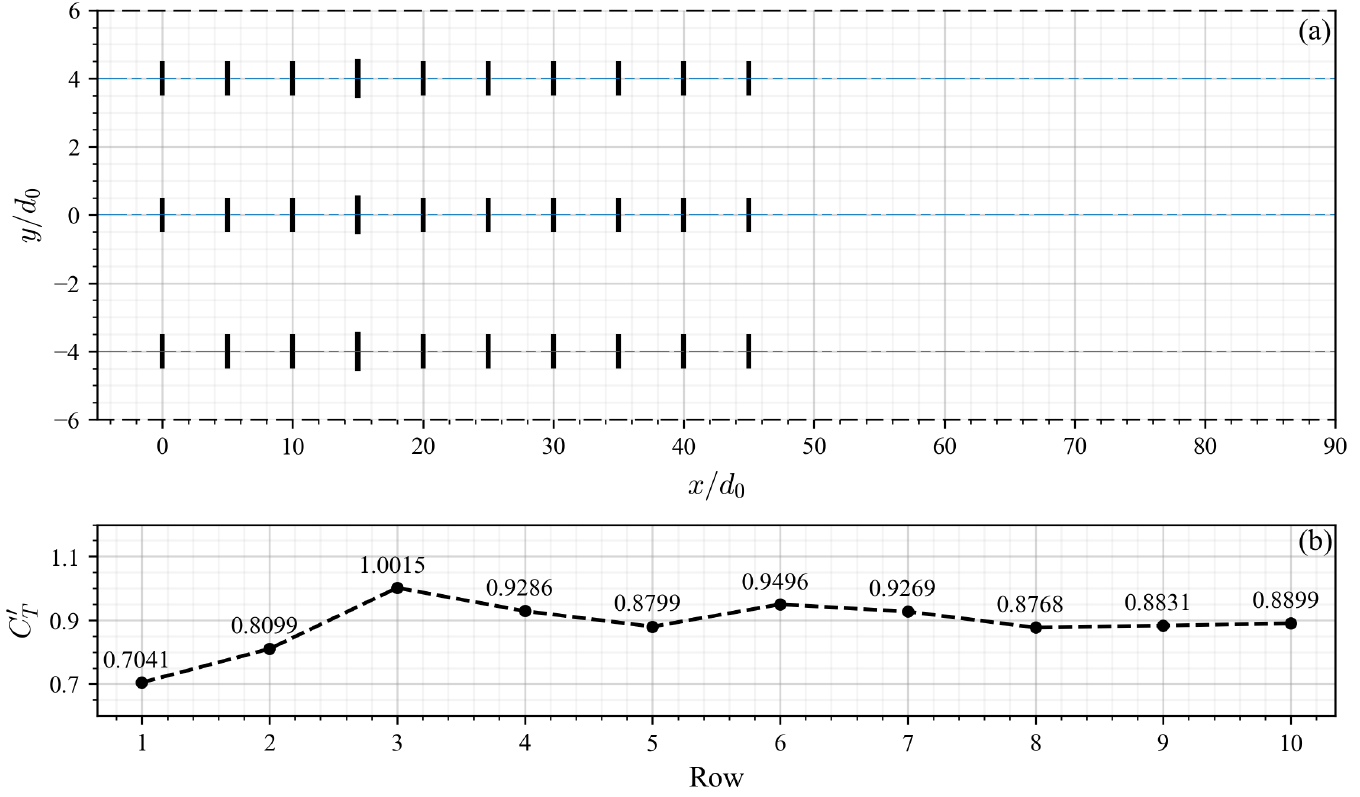}
\caption{(a) Schematic of the $10 \times 3$ wind-turbine array. Here, the black rectangles show the turbines' location. The inlet and outlet are located at $x/d_0=-5$ and 90, respectively. The black dashed lines show the cyclic boundaries. (b) Disk-based thrust coefficient ($C^\prime_T$) for different turbine rows used for RANS simulation of the $10 \times 3$ array, obtained from Ref. \cite{Stevens2018}. }
\label{fig:layout_10by3}
\end{figure}

Building upon the promising results demonstrated by the extended $k-\varepsilon$ model when applied to the validation cases, we now provide further exploration of the model’s capabilities. To this end, the wind-tunnel experiment under neutral stratification performed by Chamorro and Port{\'e}-Agel \cite{Chamorro2011} is selected as the reference. Their experimental setup consisted of a $10 \times 3$ array of miniature wind turbines with a hub height of $0.83d_0$, a streamwise spacing of $5d_0$, and spanwise spacing of $4d_0$, where $d_0 = 0.15 \text{ m}$. The reported roughness height and boundary-layer depth in the experiments were $19.98 \times 10^{-5} d_0$ and $4.5d_0$, respectively \cite{Chamorro2011,Stevens2018}. To simulate this wind farm with our RANS framework, a computational domain with a size of $95d_0 \times 12d_0 \times 4.5d_0$ in the streamwise, spanwise, and vertical directions, respectively, is created. The layout of the wind farm is schematically shown in Figure~\ref{fig:layout_10by3}(a). Similar to the validation cases, the first turbine row is placed at a distance of $5d_0$ from the domain inlet. A grid with a resolution of $450 \times 96 \times 58$ is used in the streamwise, spanwise, and vertical directions, respectively. We refine the mesh at the turbines' location in the $x$-direction and the domain is uniformly divided in the $y$-direction. Additionally, we utilize a stretching grid in the $z$-direction. Our assessment of grid independence demonstrates that this resolution effectively captures the key features of turbine wakes. The inflow characteristics match those from the experiment, and the boundary conditions are similar to the ones described in section~\ref{sec:numSetup}. As depicted in Figure~\ref{fig:layout_10by3}(b), to simulate the wind farm, we use the values of $C^\prime_T$ for the different turbine rows reported in Ref. \cite{Stevens2018}.  

\begin{figure}[!ht]
	\centering
	\includegraphics[scale=1]{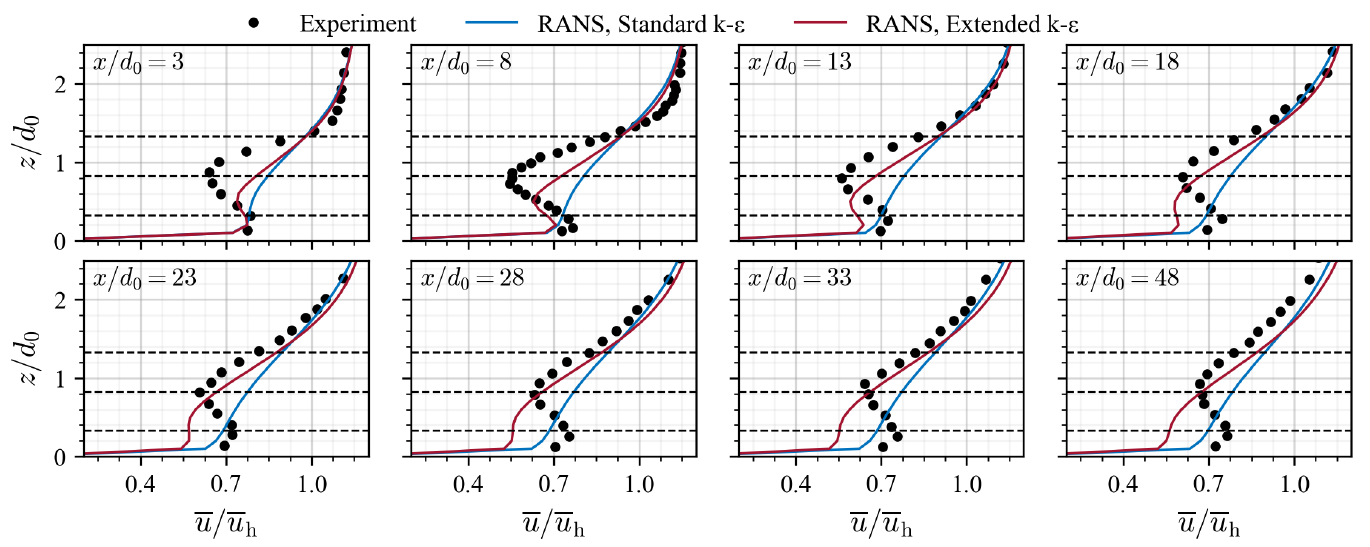}
	\caption{Vertical profiles of the normalized velocity at a distance of $3d_0$ downstream of rows 1-7 and 10 in the $10 \times 3$ array of wind turbines, predicted by the standard $k-\varepsilon$ and extended $k-\varepsilon$, and compared against data from wind-tunnel measurements. The wind-tunnel measurement profiles (black circles), which relate to the experiments performed by Chamorro and Port{\'e}-Agel \cite{Chamorro2011}, are extracted from Ref. \cite{Stevens2018}. Horizontal black dashed lines indicate the bottom-tip, hub, and top-tip heights.}
	\label{fig:vertical_vs_experiment}
\end{figure}

Figure~\ref{fig:vertical_vs_experiment} shows the vertical distributions of the normalized streamwise velocity obtained from the standard and extended $k-\varepsilon$ models compared against wind-tunnel measurement data \cite{Chamorro2011}. The profiles are extracted at a distance of $3d_0$ behind the turbine rows 1-7 and 10 \cite{Stevens2018}.  Focusing on the predictions of the standard $k-\varepsilon$ model, one can see that it generally overestimates the normalized streamwise velocity in the area from bottom- to top-tip heights, and this agrees with the earlier observations. In contrast to the standard $k-\varepsilon$ model, there is a satisfactory agreement between the results of the extended  $k-\varepsilon$ and measurement data 
within the bottom-tip to the top-tip range. This observation is also in agreement with previous findings showing that the extended  $k-\varepsilon$ model predicts more accurate disk-averaged velocities and normalized power of turbines. 

\subsection{Applying the extended $k-\varepsilon$ model to the Horns Rev 1 (HR1) wind farm}

To evaluate the performance of the extended $k-\varepsilon$ model on operational wind farms, we apply it to the HR1 wind farm and compare the obtained normalized power with LES data from Ref. \cite{Wu2015} for a hub-height inflow velocity of 8 m/s and ambient turbulence intensity of 7.7\%. The HR1 wind farm holds eighty Vestas V-80 2 MW turbines with a hub height of 70 m and a rotor diameter of 80 m. The spacing between the consecutive turbines for wind direction of 270$^\circ$, 222$^\circ$, and 312$^\circ$ are 7$d_0$, 9.3$d_0$, and 10.4$d_0$, respectively. Therefore, among these full-wake conditions, the most severe wake effect would be observed for a wind direction of 270$^\circ$. We simulate the entire wind farm with our RANS framework coupled with the standard and extended $k-\varepsilon$ models. For the sake of brevity, we omit more information on the computational domain and boundary conditions. Readers may refer to Ref. \cite{zehtab2022_2}, for more information on the methodology adopted for the simulations. 

\begin{figure}[!ht]
	\centering
	\includegraphics[scale=1]{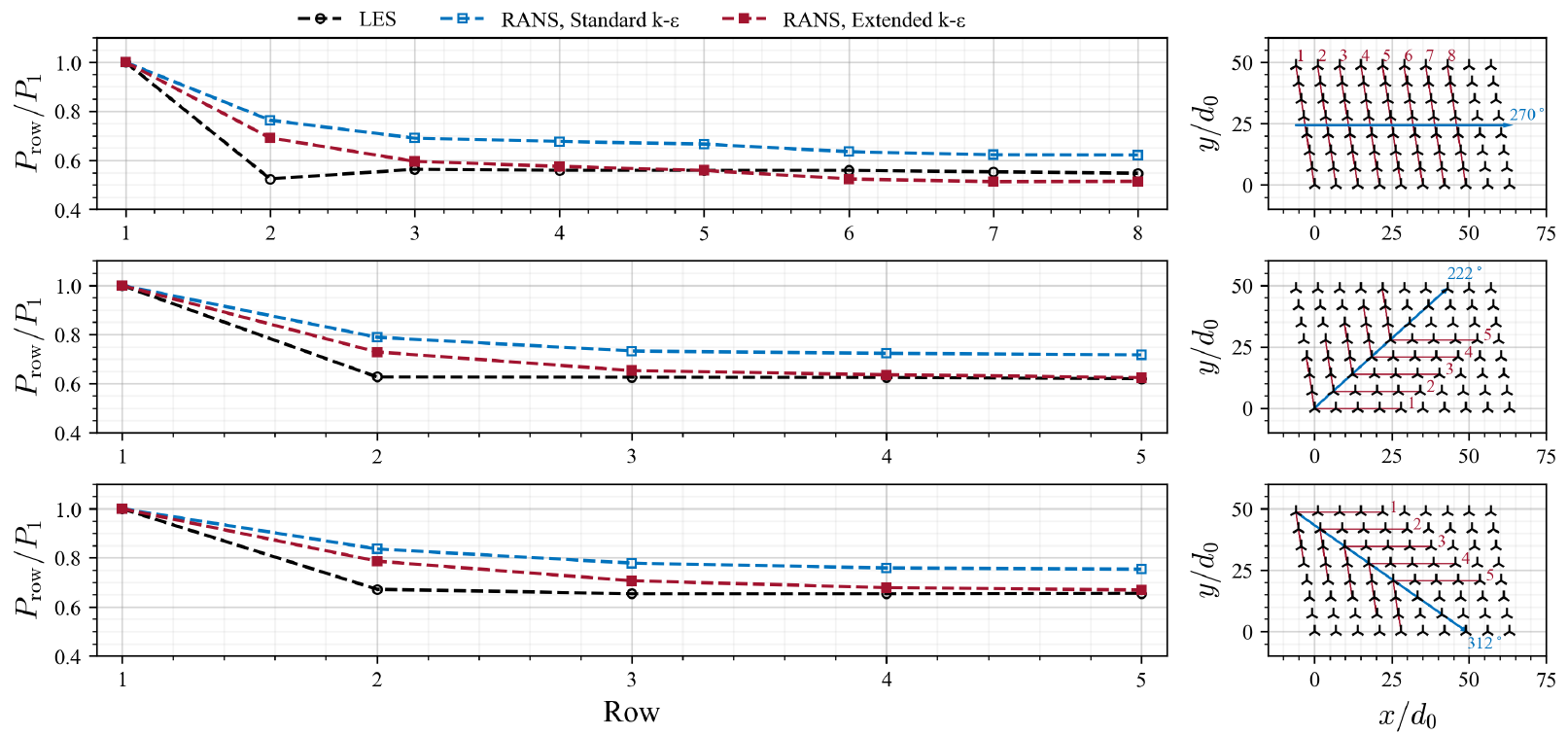}
	\caption{Left: The normalized power of turbine rows predicted by the standard and extended $k-\varepsilon$ models are compared with LES results of Wu and Port\'e-agel \cite{Wu2015} for a hub-height inflow velocity of 8 m/s and ambient turbulence intensity of 7.7\% at three different wind directions. Right: The turbines highlighted with the red lines are used for the calculation of the average power of each row, and the average value of the first row is used as the reference to calculate the normalized power.}
	\label{fig:NP_HR1}
\end{figure}

As shown in Figure~\ref{fig:NP_HR1}, the extended $k-\varepsilon$ model compares well with the LESs of Wu and Port\'e-agel \cite{Wu2015} for all three incoming wind directions. Summarized in Table~\ref{tab:HR1_summary}, one can compare the average normalized power of the waked turbine rows calculated through the standard and extended $k-\varepsilon$ models with the LES data from Ref. \cite{Wu2015}. 
Due to the elevated inflow turbulence level and larger inter-turbine spacing compared to the validation cases which translates into a milder wake, the RANS models exhibit a better performance here. 
As expected, the maximum error reduction corresponds to a wind direction of 270$^\circ$. Interestingly, the relative errors of the extended $k-\varepsilon$ model compared to the LES data are below 8\% for all three incoming wind directions.

\begin{table}[!ht]
\centering
\caption{Average normalized power of the waked turbine rows in HR1 wind farm predicted by LES \cite{Wu2015}, standard $k-\varepsilon$, and extended $k-\varepsilon$ for a hub-height velocity of 8 m/s and ambient turbulence intensity of 7.7\%.}
\label{tab:HR1_summary}
\begin{tabular}{@{}c|ccc|cc|c@{}}
\toprule
Wind    &	 \multicolumn{3}{c|}{Average of the normalized power of waked turbines}    &	\multicolumn{2}{c|}{Error}          &   Error\\
direction      &	LES	             &	Standard $k-\varepsilon$   &	Extended $k-\varepsilon$ & Standard $k-\varepsilon$  &	Extended $k-\varepsilon $          &  reduction  \\ \midrule
270$^\circ$	   &	0.55	     &	0.67	 & 0.57 &         	21.82\%		       &	3.64\%		&    18.18\%	     \\	
222$^\circ$   &	0.62	     &	0.74	  & 0.66&         	19.35\%			       &	6.45\%		 &   12.90\%	     \\	  
312$^\circ$	   &	0.66	     &	0.78	  &0.71&         	18.18\%			       &	7.58\%		  &  10.60\%	    \\			\bottomrule
\end{tabular}
\end{table}

\section{Conclusions} \label{Sec:Conclusions}
The present study aimed to propose an extended $k-\varepsilon$ model for the simulation of wakes and power losses in wind farms. It is known that the inherent shortcomings associated with the widely-utilized standard $k-\varepsilon$ model lead to overestimation of \blue eddy viscosity and \black turbulence intensity in the wake region and, as a result, an overprediction of the normalized power of the waked turbines. To solve this issue, we included an additional term resembling the impact of the turbine forces in the transport equation of turbulent kinetic energy. The new term was derived based on physical principles through an analytical approach. To assess the performance of the proposed model, initially, we studied three validation cases with different layouts under full and partial wake conditions and compared the quantities of interest, i.e., normalized velocity deficit, turbulence intensity, and turbines' normalized power, predicted by the standard and extended $k-\varepsilon$ models against LESs performed in-house \cite{Eidi2021,Eidi2022}. Then, we modeled a wind farm consisting of a $10 \times 3$ array of turbines and evaluated the success of the extended $k-\varepsilon$ model in predicting the streamwise velocity in turbine wakes through comparison with wind-tunnel measurement from Ref. \cite{Chamorro2011}. Finally, we examined the performance of our model by applying it to the operational wind farm of HR1 and compared its predictions with LES data from Ref. \cite{Wu2015}.

\blue The investigations showed that the incorporation of the term associated with the turbine-induced forces in the transport equation of the turbulent kinetic energy managed to indirectly reduce the eddy viscosity and, consequently, the mixing, in turbine wakes, as a well-known challenge linked to the standard $k-\varepsilon$ model\black. 
The turbulence level behind the turbines, predicted by the extended $k-\varepsilon$, compared well with the LES results, and unlike the standard $k-\varepsilon$ model, our extended model could capture the double peak structure of the turbulence intensity behind the turbines. As a result, the extended model could solve the issue with the standard $k-\varepsilon$ model in terms of the underestimation of velocity deficit in turbine wakes. Then, we focused on the normalized powers obtained from the standard and extended versions of the $k-\varepsilon$ model and compared them with LES. In all three validation cases, the extended $k-\varepsilon$ model outperformed the standard $k-\varepsilon$. In the context of comparison against wind-tunnel measurements, it became evident that the normalized streamwise velocity in the wake region obtained from the extended model exhibited satisfactory agreement with measurement data across the vertical distribution from bottom- to top-tip heights. When applied to the operational wind farm of HR1, the relative error between the predictions of the extended $k-\varepsilon$ and the LES data \cite{Wu2015} were below 8\% for three wind directions leading to the full-wake condition. The comparisons and validations conducted in this study proved the high potential of the proposed model for utilization by the wind-energy community. 
Applications of the present model to other complex turbulent flow problems (e.g. hilly terrain and thermally stratified ABL) could be considered as future work to further enrich the assessment of the proposed model.

\section*{CRediT authorship contribution statement}
\textbf{Navid Zehtabiyan-Rezaie}: Data curation, Formal analysis, Software, Writing – original draft. \textbf{Mahdi Abkar}: Formal analysis, Project administration, Resources, Supervision, Writing – review \& editing.
\section*{Conflict of interest}
The authors have no conflicts to disclose.
\section*{Data availability statement}
The data that support the findings of this study are available from the corresponding author upon reasonable request.
\section*{Acknowledgment}
The authors acknowledge the financial support from the Independent Research Fund Denmark (DFF) under Grant No. 0217-00038B. The computational resources used in this study were provided by the DeiC National HPC under Project No. DeiC-AU-N2-2023009.
\bibliographystyle{elsarticle-num} 

\end{document}